\documentclass[twocolumn]{aastex631}
\hypersetup{linkcolor=blue,citecolor=blue,filecolor=cyan,urlcolor=teal}

\newcommand{\gaia}{\textit{Gaia}}

\defcitealias{Leleu2024}{L24}

\graphicspath{{./}{figures/}}

\begin{document}

\title{Discovery of a Massive Outer Companion to the TOI-178 Resonant Chain}

\newcommand{\nrc}{NRC Herzberg Astronomy and Astrophysics,
5071 West Saanich Road,
Victoria, BC, V9E 2E7, Canada}
\newcommand{\ucsd}{Department of Astronomy \& Astrophysics,  University of California, San Diego, La Jolla, CA 92093, USA}
\newcommand{\nmsu}{Department of Astronomy, New Mexico State University, 1320 Frenger Mall, Las Cruces, NM 88003, USA}
\newcommand{\uvic}{Department of Physics and Astronomy, University of Victoria,
3800 Finnerty Road, Elliott Building,
Victoria, BC, V8P 5C2, Canada}
\newcommand{\mac}{Department of Physics \& Astronomy, McMaster University, 1280 Main Street West, Hamilton, ON, L8S 4L8, Canada}
\newcommand{\caltech}{Department of Astronomy, California Institute of Technology, Pasadena, CA 91125, USA}
\newcommand{\hawaii}{Institute for Astronomy, University of Hawaii at M\=anoa, 2680 Woodlawn Drive, Honolulu, HI 96822, USA}
\newcommand{\utaustin}{Department of Astronomy, The University of Texas at Austin, Austin, TX 78712, USA}
\newcommand{\ucsc}{Department of Astronomy \& Astrophysics, University of California, Santa Cruz, CA 95064, USA}
\newcommand{\ucsb}{Department of Physics, University of California, Santa Barbara, CA 93106, USA}
\newcommand{\ucla}{Department of Earth, Planetary, and Space Sciences, University of California, Los Angeles, CA 90095, USA}

\author[0000-0002-1845-2617]{Emma V. Turtelboom} \affiliation{Department of Physics \& Astronomy, McMaster University, 1280 Main Street West, Hamilton, Ontario, Canada} 

\author[0000-0001-5684-4593]{William Thompson}
\affiliation{\nrc}

\author[0000-0001-5383-9393]{Ryan Cloutier}\affiliation{Department of Physics \& Astronomy, McMaster University, 1280 Main Street West, Hamilton, Ontario, Canada}

\author[0009-0008-1229-3230]{Alexandra Rochon}
\affiliation{Department of Physics \& Astronomy, McMaster University, 1280 Main Street West, Hamilton, Ontario, Canada} 


\author[0000-0002-5627-5471]{Charles A.~Beichman}
\affil{NASA Exoplanet Science Institute-Caltech/IPAC, Pasadena, CA 91125, USA}
\affil{Jet Propulsion Laboratory, California Institute of Technology, Pasadena, CA 91109, USA}

\author[0000-0003-1499-0191]{Trevor Campbell}
\affiliation{Department of Statistics, University of British Columbia, Vancouver, BC V6T 1Z4, Canada}

\author[0000-0003-0690-1056]{Nick Choksi}
\affiliation{Division of Geological and Planetary Science, California Institute of Technology, Pasadena, CA 91125, USA}

\author[0000-0002-6773-459X]{Doug Johnstone}
\affiliation{\nrc}
\affiliation{\uvic}

\author[0000-0003-3430-3889]{Jessica Speedie}
\affiliation{Department of Earth, Atmospheric, and Planetary Sciences, Massachusetts Institute of Technology, Cambridge, MA 02139, USA}


\author[0000-0001-9582-4261]{Dori Blakely}
\affiliation{\uvic}
\affiliation{\nrc}

\author[0000-0002-9820-1884]{Simon Blouin}
\affiliation{\uvic}

\author[0000-0001-7402-8506]{Alexandre Bouchard-Côté}
\affiliation{Department of Statistics, University of British Columbia, Vancouver, BC V6T 1Z4, Canada}

\author[0000-0003-2649-2288]{Brendan P. Bowler}
\affiliation{\ucsb}

\author[0000-0002-5741-3047]{David~R.~Ciardi}
\affil{NASA Exoplanet Science Institute-Caltech/IPAC, Pasadena, CA 91125, USA}

\author[0000-0001-5173-2947]{Clarissa R. Do \'O}
\affiliation{\caltech}

\author[0009-0007-6766-2040]{Andre Fogal}
\affiliation{\uvic}

\author[0000-0003-4557-414X]{Kyle Franson}
\affiliation{\ucsc}

\author[0009-0009-2223-2404]{Kaitlyn Hessel}
\affiliation{\uvic}

\author[0000-0002-4164-4182]{Christian Marois}
\affiliation{\nrc}
\affiliation{\uvic}

\author[0000-0001-6975-9056]{Eric Nielsen}
\affiliation{\nmsu}

\author[0009-0008-9687-1877]{William Roberson}
\affiliation{\nmsu}

\author[0000-0003-2233-4821]{Jean-Baptiste Ruffio}
\affiliation{\ucsd}

\author[0000-0002-4290-6826]{Judah Van Zandt}
\affiliation{\ucsb}

\author[0000-0002-6618-1137]{Jerry W. Xuan}
\affiliation{\ucsb}

\author[0000-0002-2696-2406]{Jingwen Zhang}
\affiliation{\ucsb}

\correspondingauthor{Emma V. Turtelboom}
\email{eturtelboom@berkeley.edu}

\begin{abstract}
We report the discovery of a massive outer companion around TOI-178, a known planetary system with six inner small planets within 0.15 AU, five of which are in a 2:4:6:9:12 resonant chain. We analyse the astrometric anomalies in the combined \gaia{} DR2/DR3 -- Hipparcos catalogue (G23H) and literature radial velocities, revealing strong evidence for a massive outer companion at several AU. We combine the resulting orbital fit posterior with archival high-contrast imaging to constrain the properties of the corresponding massive outer companion, TOI-178\,h. We also study the dynamical stability of the inner resonant chain in the presence of this previously unknown outer companion via both heuristic and N-body simulations. We find that the inner system appears to be robust to the perturbing influence of TOI-178\,h, in line with the systems several Gyr age. Our analysis results in a final posterior solution for TOI-178\,h with a semi-major axis of $6.9^{+12.3}_{-3.3}$\,AU, a mass of $21^{+47}_{-11}\,M_J$, and eccentricity of $0.76^{+0.18}_{-0.35}$, likely corresponding to an eccentric brown dwarf. TOI-178 is currently the only known system that has maintained a high-multiplicity resonant chain whilst in the dynamical presence of a massive outer companion, showcasing a unique outcome of planet formation.
\end{abstract}


\section{Introduction} \label{sec:intro}
The majority of the known population of exoplanets have close separations from their host stars; only $\sim2\%$ of the confirmed planets have measured semi-major axes beyond 5 AU\footnote{NASA Exoplanet Archive, \url{https://exoplanetarchive.ipac.caltech.edu}, accessed 22 October 2026}. This is primarily due to observational biases in the transit and radial velocity methods, which have been used to detect the vast majority of planets to date. However, complementary observational methods such as microlensing, direct imaging and astrometry are sensitive to long-period, giant planets. 

The Gaia mission collected astrometric data for $\sim 2.8$ billion stars across more than a decade \citep{gaia}. The fourth data release (Gaia DR4, expected December 2026) will provide high-quality time series astrometry for $\sim 2$ billion stars as well as a catalogue of detected companions, enabling a diverse range of scientific analyses. Astrometry has already be used alongside radial velocity data to discover a handful of planets, both from ground-based facilities \citep{Sahlmann2013} and the previous Gaia data release (Gaia DR3) of astrometric solutions \citep{stefansson2025}. All of these planets are $>2 \rm{M_J}$ and have orbital periods longer than 280 d. Gaia DR4 will likely reveal thousands of additional giant planet candidates at several AU \citep{lammerswinn2026}. Some of these candidates will orbit stars with already-known inner planets. 

TOI-178 is one such system, hosting six inner transiting planets, five of which are in a resonant chain \citep{Leleu2024}. Young multi-planet systems appear to host near-resonant planet pairs more frequently than  adolescent and mature systems \citep{Dai2024}. This suggests that planets are captured into resonant states as they form \citep{Izidoro2017}, but are then disrupted from resonance on $\sim100$ Myr timescales \citep[e.g.][]{Choksi2026, Hadden2026, Ogihara2026, LoRusso2026, Pichierri2020}. Resonant chains of planets around mature stars are thus considered to be pristine fossils of planet formation which have not yet been dynamically disrupted from resonance, and are prime windows into primordial planetary system architectures, raising questions of how these systems were able to resist disruption across their lifetimes.

One potential cause of planetary system disruption is the presence of a massive companion, ranging from giant planets to binary stars. External companions can perturb planets interior to their orbits, for example through the von Zeipel-Kozai-Lidov mechanism \citep{vonZeipel1910, Kozai1962, Lidov1962}. This perturbation can occur on a large range of physical scales from asteroids \citep{Kozai1962} to black holes \citep{Mazzolari2022}, and has been invoked to explain a variety of observational results including spin-orbit misalignments \citep{Rice2022}, the Kepler dichotomy \citep{Rodet2021}, and Hot Jupiter formation in stellar binaries \citep{Weisserman2025,Weisserman2026}. 

We describe the current understanding of the planetary system around TOI-178 in Section \ref{sec:178background}. We then present astrometric, adaptive optics (AO), and radial velocity (RV) observations of TOI-178 in Section \ref{sec:data}, and report a new outer companion (OC), TOI-178\,h, in the system based on the joint RV and astrometric posterior. We then use the AO imaging to further constrain the OC's properties in Section \ref{sec:methods}. In Section \ref{sec:stability} we use both heuristic stability criteria and N-body simulations to study the impact of TOI-178\,h on the inner resonant chain. We present our final posterior in Section \ref{sec:results}. We discuss the impacts of this companion on the system's stability and potential formation in Section \ref{sec:discussion} and conclude in \ref{sec:conclusions}.

\section{TOI-178 Inner System} \label{sec:178background}

TOI-178 is a bright K-type star ($T_{\rm eff} = 4316 \pm 70$ K, $G = 11.158 \pm 0.003$) that hosts six confirmed transiting planets \citep{leleu2021}. The star has a slightly elevated Renormalized Unit Weight Error (\texttt{ruwe}) of 1.55, which may indicate binarity, but the percentage of successful image parameter determination (IPD) windows with more than one PSF peak (\texttt{ipd\_frac\_multi\_peak}) is 0, and the star does not have a resolved companion \citep{GaiaCollaboration2023}. We summarize the stellar parameters of TOI-178 in Table~\ref{tab:star}.

\begin{table}
\centering
\small
\label{tab:star}
\caption{Stellar parameters for TOI-178 sourced from [1] \cite{tic}, [2] \cite{GaiaCollaboration2023}, [3] \cite{2mass}, [4] \cite{delrez2023} [5] \cite{leleu2021} }
\begin{tabular}{lccc}
\hline\hline
ID &  &  & Source \\
\hline
TIC & 251848941 &  & [1] \\
Gaia DR3 & \multicolumn{2}{l}{2318295979126499200} & [2]\\
\hline\hline
Parameter & Value & Uncertainty & Source \\
\hline
G magnitude & 11.1575 & 0.002 & [2] \\ 
J magnitude & 9.372 & 00.021 & [3] \\
K magnitude & 8.656 & 0.021 & [3] \\
Parallax (mas) & 15.900 & 0.031 & [2] \\
Distance (pc) & 62.89 & 0.12 & from parallax \\
RUWE & 1.549344 & -  & [2] \\
Age (Gyr) & 6.0 & $^{+6.8}_{-5.0}$ & [4] \\
$\rm{M_* (M_\odot)}$ & 0.647 & $^{+0.030}_{-0.029}$  & [4] \\
$\rm{T_{eff} (K)}$ & 4316 & 70  & [5] \\
$\rm{[Fe/H]}$ (dex) & -0.23 &  0.05 & [5] \\
\hline
\end{tabular}
\end{table}

The six planets orbiting TOI-178 were initially confirmed using transit and radial velocity observations, namely TESS\footnote{Transiting Exoplanet Survey Satellite,  \citet{Ricker2015}} (Sector 2), CHEOPS\footnote{ CHaracterising ExOPlanet Satellite, \citet{Benz2021}}, VLT/ESPRESSO\footnote{Echelle SPectrograph for Rocky Exoplanets and Stable Spectroscopic Observation, \citet{Pepe2021}}, NGTS\footnote{Next Generation Transit Survey, \citet{Wheatley2018}}, and SPECULOOS\footnote{Search for habitable Planets EClipsing ULtra-cOOl Stars, \citet{Sebastian2021}} data \citep{leleu2021}. A follow-up work performed a photodynamical analysis of the system using this data as well as additional CHEOPS, NGTS, and EulerCam observations \citep[][hereafter \citetalias{Leleu2024}]{Leleu2024}. \citetalias{Leleu2024} found that the five planets c to g are locked in a resonant chain with period ratios of 2:4:6:9:12 and librating Laplace angles. The innermost planet, TOI-178 b, is disconnected from this chain. This result places TOI-178 among a very small sample of seven high-multiplicity ($N_{pl} > 4$) planetary systems with planets formally in resonant states, making the system a unique arena to study the formation and evolution of resonant chains. 

\begin{table*}
\centering
\small
\label{tab:planets}
\begin{tabular}{ccccc}
\hline\hline
Planet & $P$ & $P_{i}/P_{i-1}$ & $R_p$ & $M_p$ \\
 & (d) &  & ($R_\oplus$) & ($M_\oplus$) \\
\hline
b* & $1.914560 \pm 0.000004$ & - & $1.200 \pm 0.039$ & $0.96 \pm 0.68$ \\
c & $3.238486 \pm 0.000012$  & 1.69 & $1.754 \pm 0.036$ & $4.64 \pm 0.52$ \\
d & $6.557569 \pm 0.000060$  & 2.02 & $2.695 \pm 0.044$ & $5.20 \pm 0.41$ \\
e & $9.963180 \pm 0.000135$  & 1.52 & $2.301 \pm 0.038$ & $3.48 \pm 0.29$ \\
f & $15.233350 \pm 0.000245$ & 1.53 & $2.417 \pm 0.044$ & $5.63 \pm 0.43$ \\
g & $20.716630 \pm 0.000305$ & 1.36 & $2.939 \pm 0.056$ & $4.40 \pm 0.38$ \\
\hline
\end{tabular}
\caption{Planet parameters reported in \citetalias{Leleu2024} \newline * TOI-178 b is not formally in resonance with the five other planets, which form a 2:4:6:9:12 resonant chain.}
\end{table*}

\begin{figure}[t!]
    \centering
    \includegraphics[width=\columnwidth]{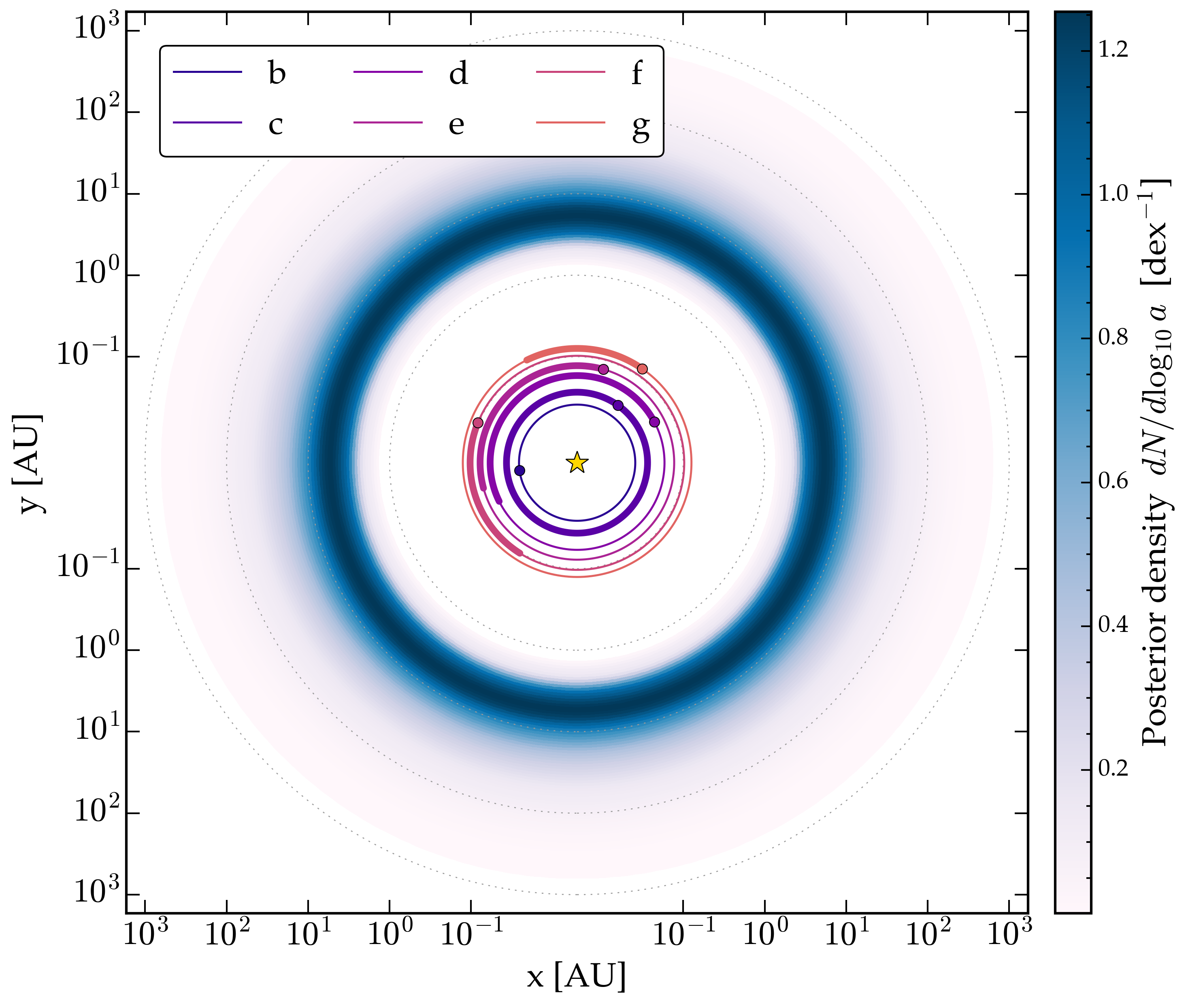}
    \caption{Representation of the TOI-178 system. Orbits for the inner planets are drawn from the posteriors of \citetalias{Leleu2024}. Fractions of the inner planets' orbits are shaded relative to the orbital period of TOI-178 c, the innermost planet in the resonant chain. Decades in semi-major-axis are shown as concentric dashed grey lines. The outer companion's final marginalized semi-major-axis posterior range (described in Section \ref{sec:results}) is highlighted in blue and shaded according to posterior density in log-semi-major-axis space. We note that this does not depict orbit draws of the OC, which is likely eccentric. Rather, we depict only the range of semi-major axes in the final OC posterior.}
    \label{fig:orbits}
\end{figure}

The properties of the inner transiting planets are summarized in Table~\ref{tab:planets}. The TOI-178 planets have orbital periods between $\sim 1.9 - 20.7$ days, with radii between $1.2 -2.9\, \rm{R_\oplus}$, and masses ranging from $\sim 1 - 5.6\, \rm{M_\oplus}$. The innermost planet TOI-178\,b is roughly Earth-mass and is $\sim20\%$ larger than the Earth. The remaining five planets are sub-Neptunes, with an uncharacteristically broad intra-system range of bulk densities: planet c has a bulk density of $4.8 \pm 0.6 \rm{g/cm^3}$ while planet g has a bulk density of only $1.0 \pm 0.1 \rm{g/cm^3}$. This spread suggests a wide range of bulk compositions possibly reflecting unique formation environments or evolutionary pathways prior to being captured into resonance. The system's orbital architecture in the presence of the outer companion TOI-178\,h is depicted in Fig.~\ref{fig:orbits}.

\section{Data}\label{sec:data}

In this section we describe the observational data used in our analysis. Specifically, we detail the astrometric (Gaia and Hipparcos) and RV (VLT/ESPRESSO) data which we jointly fit to reveal an OC around TOI-178. We also introduce AO (Keck-II/NIRC2) imaging which we used to further constrain the parameters of TOI-178\,h.

\subsection{Astrometry}

The primary data used to discover the OC TOI-178\,h is the calibrated, absolute astrometry from the G23H catalogue \citep{Thompson2026}. The G23H catalogue presents posteriors from astrometric orbital fits using \texttt{Octofitter} \citep{octofitter} for a large sample of stars with astrometric measurements from both the \textit{Hipparcos} \citep[1989-1993;][]{hipparcos} and \textit{Gaia} \citep[2014-2025;][]{gaia} missions. Specifically, G23H contains Gaia DR2 - DR3 proper motion anomalies \citep{Thompson2026, Thompson2026b}, Hipparcos - Gaia DR3 proper motion anomalies \citep{Brandt2021}, calibrated Gaia DR3 Renormalized Unit Weight Error (RUWE) constraints \citep{Kiefer2025} and Gaia DR3 Radial Velocity Spectrograph excess noise \citep{Chance2025}. This rich astrometric catalogue has been used to constrain the masses of perturbers in transition disks \citep{blakely2026}, study the brown dwarf desert \citep{bddesert}, search for outer giant planets in known planetary systems \citep{doo2026}, and measure the conditional occurrence of outer companions given inner small planets \citep{rochon2026}.  

\subsection{Radial Velocities} \label{sec:rvs}

\citet{leleu2021} presented 46 RVs collected of TOI-178 using VLT/ESPRESSO across 112 days (29 September 2019 to 20 January 2020). The median error of the RVs is 0.73 m/s. These data place additional constraints on the OC around TOI-178, as the RVs show little scatter (maximum RV variation of $\approx$ 25 m/s), nor evidence of a long-term trend or offset due to a distant massive companion. The RV data alone do not show clear evidence for the OC. We jointly fit the astrometric and median-subtracted RV data as described in Section~\ref{sec:methods}.

\subsection{Adaptive Optics Imaging}
There are several imaging observations of TOI-178 reported on ExoFOP using both AO and speckle imaging\footnote{\url{https://exofop.ipac.caltech.edu/tess/target.php?toi=178}, accessed 7 Nov 2026}. 
In this analysis, we use Keck-II/NIRC2 observations as they are sensitive to the smallest angular separations of the available imaging data. TOI-178 was observed by the NIRC2 instrument on Keck-II \citep{Wizinowich2000} on 2018-Nov-22 UT (specifically beginning at 2458444.23332 BJD) in a standard 3-point dither pattern. The dither pattern step size was $3\arcsec$ and was repeated twice, with each dither offset from the previous dither by $0.5\arcsec$ for a total of 9 images. NIRC2 was used in the narrow-angle mode with a full field of view of $\sim10\arcsec$ and a pixel scale of approximately $0.0099442\arcsec$ per pixel. The observations were made in the narrow-band filters Br$\gamma$ filter $(\lambda_o = 2.1686; \Delta\lambda = 0.0326~\mu$m) with an integration time in each filter of 4s for a total of 36s on target.

The science frames were flat-fielded and sky-subtracted. The flat fields were generated from a median average of dark subtracted flats taken on-sky.  The flats were normalized such that the median value of the flats is unity. The sky frames were generated from the median average of the 15 dithered science frames; each science image was then sky-subtracted and flat-fielded. The reduced science frames were combined into a single combined image using an intra-pixel interpolation that conserves flux, shifts the individual dithered frames by the appropriate fractional pixels, and median-coadds the frames. The final resolution of the combined dithers was determined from the full-width half-maximum of the point spread function; 0.067\arcsec.

\begin{figure}[t!]
    \centering
    \includegraphics[width=\linewidth]{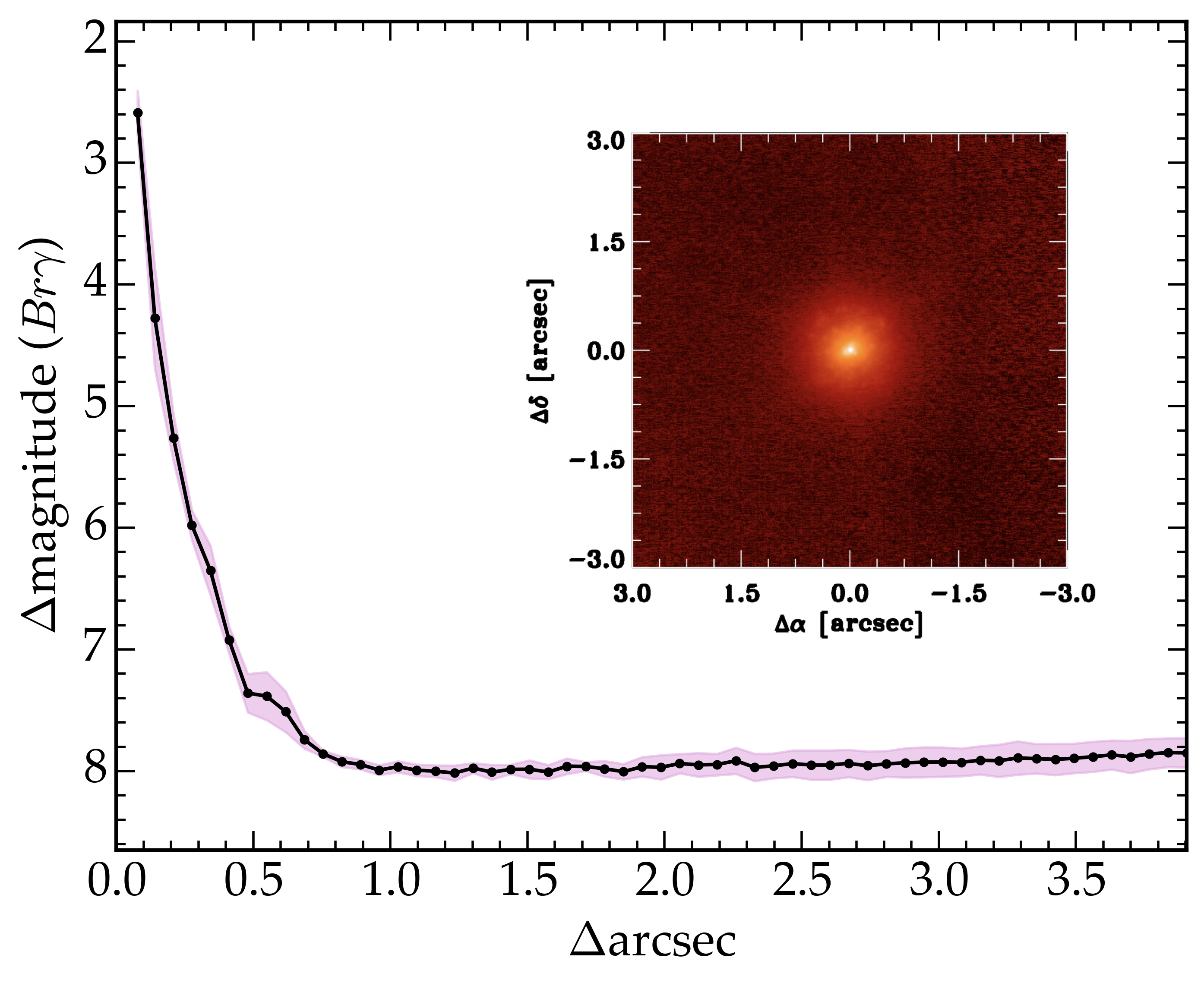}
    \caption{Contrast curve for TOI-178 taken using the Br$\gamma$ filter on Keck/NIRC2. \textit{Inset: } AO image of TOI-178. These observations did not identify any companions.}
    \label{fig:aoimage}
\end{figure}

Within the limits of the AO observations, no stellar companions were detected. The sensitivities of the combined image were determined by injecting simulated sources azimuthally around the primary target every $20^\circ $ at separations of integer multiples of the central source's FWHM \citep{Furlan2017}. The brightness of each injected source was scaled until standard aperture photometry detected it with $5\sigma $ significance. The resulting brightness of the injected sources relative to TOI~178 set the contrast limits at that injection location. The final $5\sigma $ limit at each separation was determined from the average of all of the determined limits at that separation and the uncertainty on the limit was set by the root mean square dispersion of the azimuthal slices at a given radial distance (Fig.~\ref{fig:aoimage}).

\section{Methods} \label{sec:methods}

\subsection{Joint Astrometry \& RV Fit}
We jointly fit the G23H astrometric and RV data using the \texttt{Octofitter} model framework \citep{octofitter}. The six inner planets around TOI-178 have signals far below the astrometric detection limit of the G23H catalogue, and thus are not detectable in this fit \citep{Thompson2026, Thompson2026b}. We implemented a similar fit as in \citet{Thompson2026}, but also incorporate the ESPRESSO RVs. The \texttt{Octofitter} framework performs Bayesian modelling of the G23H catalogue quantities, combining the log-likelihood functions of a model given the Gaia astrometric excess noise, Gaia proper-motion epochs, and the Gaia RV variability. It also includes a log-likelihood function for the RV data set. We set broad priors on the RV offset ($\mathcal{U}(-50, 50)$\,km/s) and RV jitter ($\log{\mathcal{U}}(0.1, 100)$\,m/s). We set a stricter prior on the RV linear slope ($\mathcal{U}(-0.1,0.1)$\,m/s/d) so that large RV trends due to e.g. a binary companion would not be cancelled out by an artificially inflated RV slope. \texttt{Octofitter} also incorporates the light contributed by luminous companions when modelling the astrometric motion. 

\begin{figure}[t!]
    \centering
    \includegraphics[width=\columnwidth]{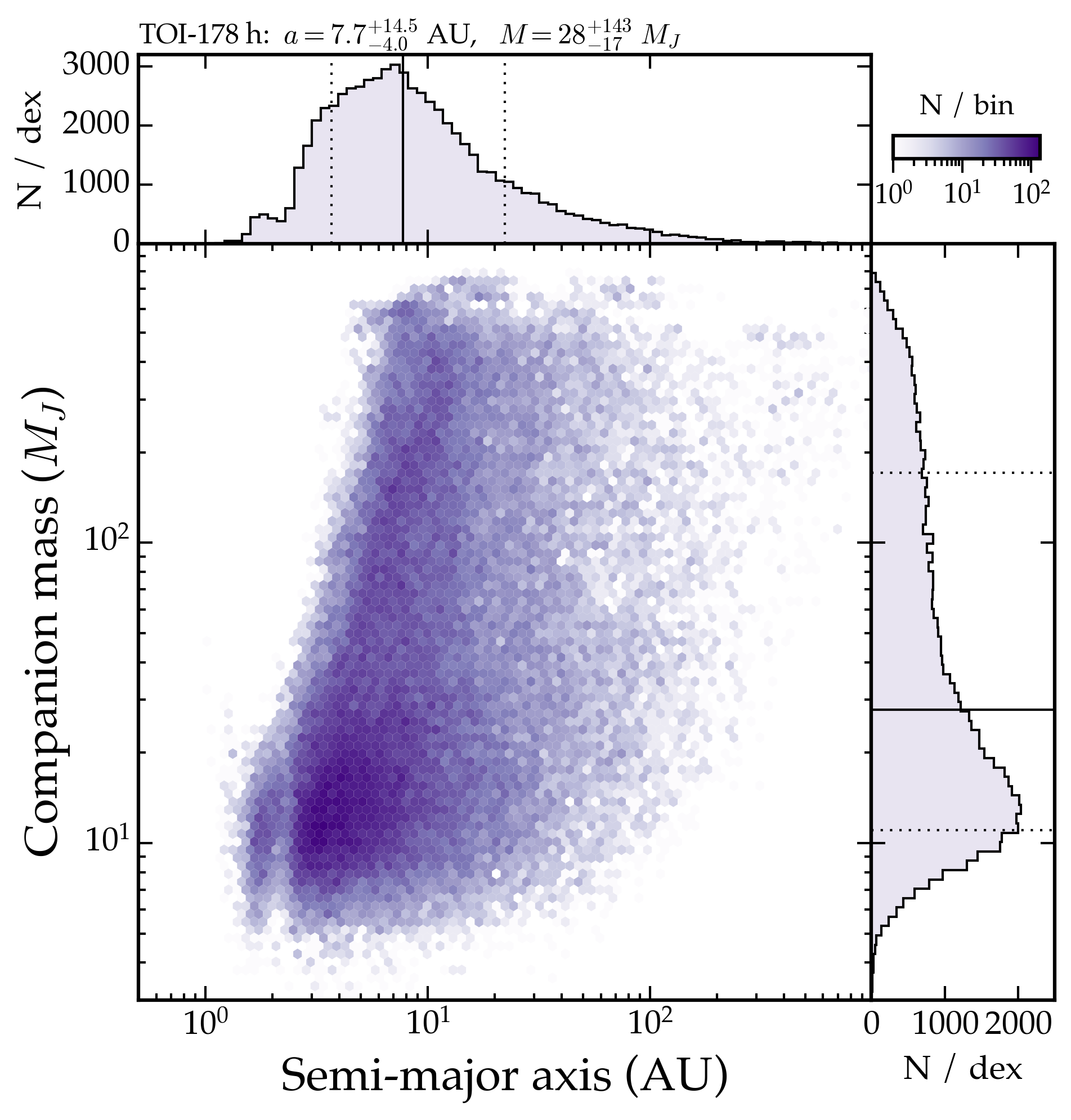}
    \caption{Binned posterior draws of the combined G23H astrometry and the RV data for TOI-178. The colour map indicates the density of posterior draws in a given hexagonal bin in semi-major axis and mass space. Marginalized histograms are shown to the top and right of the main plot, with a solid line indicating the median value and dashed lines at the 16th and 84th percentiles of the marginalized distributions.}
    \label{fig:initialpost}
\end{figure}

We obtained full orbital posteriors and Bayes factors, allowing us to identify the single-companion model as preferred given the G23H and RV data for TOI-178. Fig.~\ref{fig:initialpost} shows the binned posterior distribution of 65,536 draws. We note the notch-like feature at $\sim2$ AU; this is due to aliasing in the astrometric fit. The Gaia DR2/DR3 and Hipparcos observational baselines are $\sim 5$ years, which corresponds to $\sim 2$ AU around TOI-178; companions near this semi-major axis are thus under-detected in G23H data \citep{Thompson2026}.

\subsection{Adaptive Optics Imaging Constraints} \label{sec:ao}

We used the Keck-II/NIRC2 AO imaging described in Section~\ref{sec:ao} to further constrain the presence of self-luminous companions to TOI-178. These data are independent of our astrometric and RV data and provide complementary constraints on the outer companion. For each draw from our G23H+RV posterior, we calculated the projected physical separation, $s$, of the corresponding OC from TOI-178 at the AO imaging epoch. We then compared this projected separation to the separation limits of our Br$\gamma$ contrast curve (see Fig.~\ref{fig:aoimage}), which are between 0.078-3.904\arcsec, corresponding to projected physical separations of $\sim 5-240$ AU. Out of the 65,536 posterior $s$ draws, 46,860 (i.e. 72\%) were not observable at the AO imaging epoch and thus are not ruled out by AO imaging even if the nominal OC mass and separation of those draws lie above our AO contrast curve. 

For the posterior draws that had projected angular separations $\rho >$ 0.078\arcsec\ at the imaging epoch, we calculated the expected magnitude of the companion. We used the \citet{baraffe2015} evolutionary models (henceforth BHAC15) to calculate the predicted contrast of a given posterior draw. We did not apply this methodology to 18,676 draws which would appear interior to the innermost separation of the AO contrast curve. We also did not apply this methodology to the 46,169 ($\sim70\%$) posterior draws with companion masses $\lesssim 0.07 M_\odot \approx 75 M_J$, as these very low-mass stars or brown dwarfs have quite distinct evolutionary tracks \citep[e.g.][]{phillips2020}. In these cases, the BHAC15 models are not appropriate to calculate their expected magnitudes. 

The BHAC15 models depend on stellar age, and we used the isochrone age associated with each \texttt{Octofitter} posterior draw \citep{octofitter}. We found that the resulting magnitudes are largely independent of system age as TOI-178, and its companion if stellar, are firmly on the main sequence. We interpolated the BHAC15 models linearly in stellar mass to obtain the predicted magnitude of a companion posterior draw. We then calculated the predicted flux of the companion using the parallax returned by \texttt{Octofitter}, and subtracted this predicted companion flux from the total flux calculated from the 2MASS $K_s$-band magnitude for TOI-178 ($K_s=8.656 \pm 0.021$). We assumed that the 2MASS magnitude is the blended magnitude for TOI-178 and a putative companion due to the 2MASS pixel scale of 2.0\arcsec \citep{2mass}. We obtained a flux, and thus magnitude, for TOI-178 alone, which we used to calculate the predicted contrast between TOI-178 and the companion for each draw. We compared the resulting contrast with the interpolated AO contrast curve. 

\begin{figure*}[h!]
    \centering
    \includegraphics[width=\linewidth]{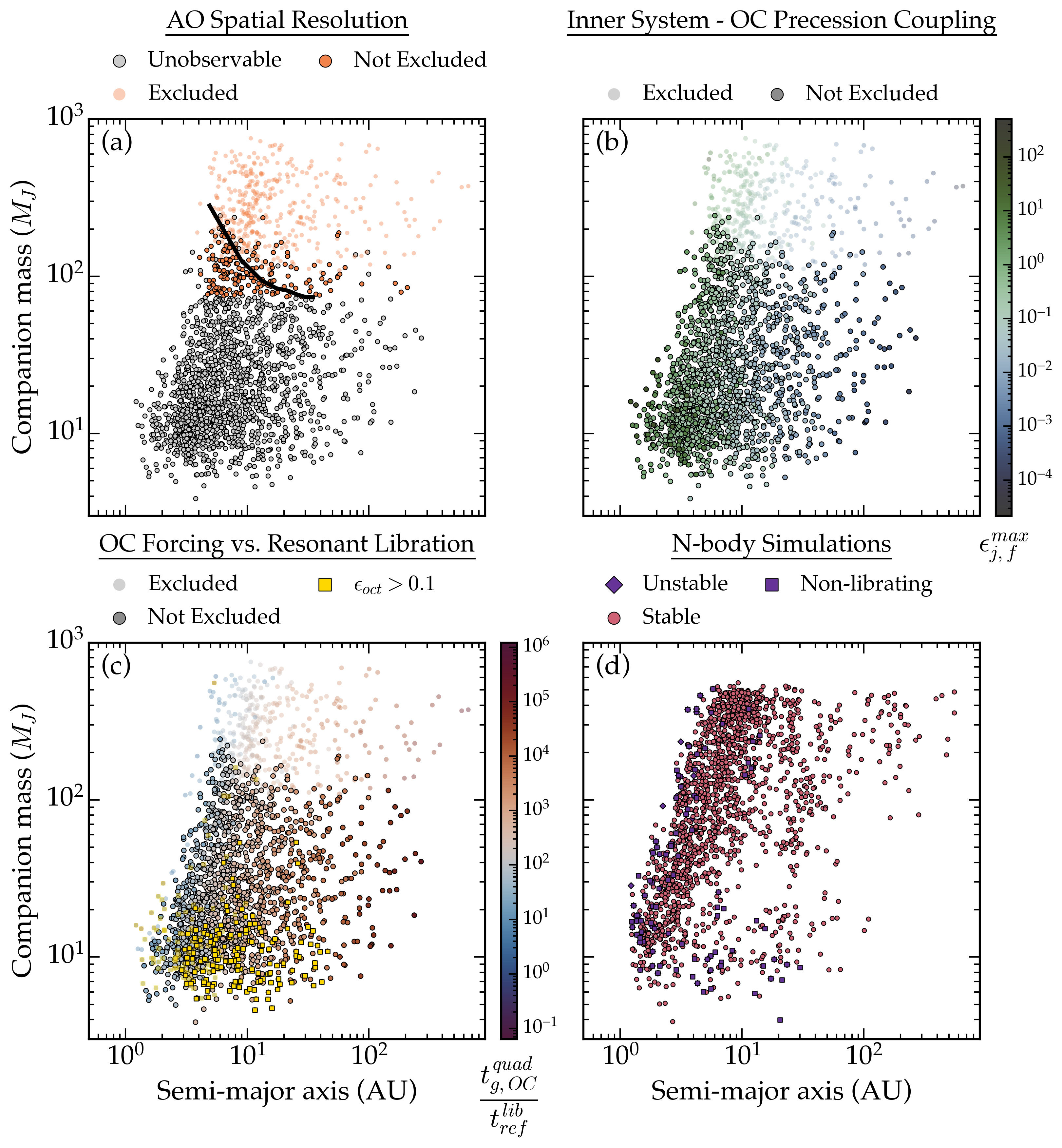}
    \caption{Random subsample of 3000 G23H+RV posterior draws (for visual clarity) for the OC TOI-178\,h in semi-major axis--mass space. In each panel, draws ruled out by a preceding constraint are shown faded, and those still allowed are outlined in black. \textbf{(a) AO Spatial Resolution:} Grey markers are posterior draws for which the imaging data give no constraint. Red points are draws tested against the AO contrast curves; faded red points are excluded by imaging or photometry. The solid line shows the Br$\gamma$ mass limit at 6\,Gyr for reference. \textbf{(b) Inner System Coupling:} Secular coupling of the inner planets relative to each OC posterior draw following \citet{Lai2017}. Draws in the weak regime ($\epsilon_{j,f}^{max} > 10$) would rapidly destabilise the inner system. \textbf{(c) OC Libration Forcing:} The colour map shows the ratio of the OC's quadrupole forcing timescale, $t_{g,\mathrm{OC}}^\mathrm{quad}$, to the longest libration timescale of the inner resonant chain, $t_\mathrm{ref}^\mathrm{lib}$. Samples with ratio $< 10$ are excluded. Gold squares mark draws where the quadrupole approximation is not valid. \textbf{(d) N-body Simulations:} Outcome of N-body integrations of the inner system plus OC over 60 $t_{g,\mathrm{OC}}^\mathrm{quad}$; purple diamonds represent simulations that became unstable over the integration time, and purple squares represent simulations that remain stable but do not librate at the end of the integration time.}
    \label{fig:theplot}
\end{figure*}

Posterior draws that lie above our observed contrast curve in panel a) of Fig.~\ref{fig:theplot} were excluded, as we would expect to have detected the corresponding companion if it were to exist. In total, we tested 18,676 posterior draws and excluded 12,176 of them using our Keck/NIRC2 AO imaging. Ruled out OCs were primarily massive ($\gtrsim 10^2 \rm{M_J}$) companions, as shown in panel a) of Fig.~\ref{fig:theplot}.

\section{Stability Analyses} \label{sec:stability}

The joint astrometry-RV posterior of the OC spans two orders of magnitude in both mass and semi-major axis. By comparing the posterior with AO imaging, we restricted the parameter space of TOI-178\,h. Here, we performed additional dynamical tests into order to further constrain the OC's parameters.

\subsection{Coupling of the inner resonant chain} \label{sec:stab-eps}

For draws with semi-major axes $\rm{a_{\rm OC}} \gg \rm{a_g}$, the system can be modelled using a hierarchical three-body secular approximation \citep[][and references therein]{Naoz2016}. We thus classified the impact of the OC on the inner system using the coupling coefficient $\epsilon_{ij}$. This coefficient describes the relative coupling between two inner planets and an outer perturber following \citet{Lai2017},
\begin{equation}
    \epsilon_{jf} \approx \frac{M_{\rm OC}}{m_f} \left(\frac{a_f}{a_{\rm OC}}\right)^3 \left(\frac{3a_j/a_f}{b_{3/2}^{(1)}(a_j/a_f)}\right) \frac{(a_f/a_j)^\frac{3}{2} - 1}{1+(L_j/L_f)}
\end{equation}
where $M_{\rm OC}$ is the mass of the outer companion, $m_f$ is the mass of planet f, $a_f$ is the semi-major axis of planet f, $a_{\rm OC}$ is the semi-major axis of the OC, $a_j$ is the semi-major axis of a given planet j (with j including planets b to e), $b_{3/2}^{(1)}$ is the Laplace coefficient \citep[Equation 5 in ][]{Lai2017}, and $L_j$ and $L_f$ are the angular momenta of a given planet j and planet f, respectively. We calculate the coefficient for each planet relative to TOI-178\,f, as planet f dominates the mass and orbital angular momentum budget of the inner system. The remaining planets in the system are steered by planet f in their response to the outer companion's perturbation. We thus calculated the maximum coupling coefficient between any of the inner planets and planet $f$ for each posterior draw. 

The posterior draws can thus be classified into three regimes: strongly-coupled ($\epsilon_{jf} \ll 1$), weakly-coupled ($\epsilon_{jf} \gg 1$), and ambiguous ($\epsilon_{jf} \sim 1$). In the strong coupling regime, the inner planets precess as a single plane, without significantly perturbing their initial mutual inclinations. In the weak coupling regime, the inner planets' orbital angular momentum vectors precess independently around that of the outer companion. This can lead to large mutual inclinations among the inner planets, which may result in a subset of the inner planets ceasing to transit along our line of sight. We consider posterior draws with $\epsilon_{jf} > 10$ to be weakly-coupled (505/65536 $\sim$ 1\%) and thus incompatible with the observed architecture of TOI-178's inner system.

In panel b) of Fig.~\ref{fig:theplot}, we show the posterior shaded by $\epsilon_{jf}$. The inner system is ambiguously coupled for the majority of the OC draws, motivating further analysis of the chain's resilience to TOI-178\,h. We note that this prescription does not consider the impact of resonance on the response of inner planets to external companions. \citet{Rodet2021} find that second-order MMRs can stiffen multi-planet systems, reducing the impact of an external perturbation on mutual inclinations within the inner system. Therefore, the results presented in panel b) of Fig.~\ref{fig:theplot} are conservative and likely underestimate the resonant chain's resilience to the OC TOI-178\,h.

\subsection{Resonant chain disruption timescales} \label{sec:stab-times}

We also estimate the de-stabilizing impact of the OC by comparing relevant timescales, described below in order of decreasing magnitude.
\begin{enumerate}
    \item The forcing timescale of the OC on the inner system under the quadrupole approximation of the Hamiltonian is 
    \begin{equation}
        t_{OC, g}^{\rm quad} \sim \frac{16}{30\pi} \frac{M_* + m_{g} + M_{\rm OC}}{M_{\rm OC}} \frac{{P_{\rm OC}}^2}{P_{g}} (1-{e_{\rm OC}}^2)^\frac{3}{2}
    \end{equation}
    following \citet{Antognini2015}. We include only planet g of the inner planets as it experiences the strongest forcing by the outer companion and thus minimises $t^{\rm quad}$. The forcing timescale of the outer companion on planet g spans eight orders of magnitude [10$^{-1}$, 10$^7$ yr] across the full G23H+RV posterior.  
    \item The libration timescale of the Laplace resonant angles for the resonant chain between planets c and g.  We measured the libration periods of the three Laplace angles using an N-body \texttt{REBOUND} integration of the adopted inner-system sample without a companion \citep{rebound1}. We ran the integration for 500 yr with the \texttt{WHFast} integrator \citep{rein2015} and general relativity included via \texttt{REBOUNDx} \citep{reboundx}. The Laplace angles ($\Psi_1, \Psi_2, \Psi_3$) were computed from the Jacobi mean longitudes every 0.02 yr. Each Laplace angle remained bounded throughout, confirming that the adopted sample lies within the librating chain \citepalias[][see their Fig. 8]{Leleu2024}. We measured each angle’s libration period from the dominant peak of a Fourier transform of its unwrapped time series, excluding the zero-frequency term. This gives $\sim$ 56, 13 and 28 yr for $\Psi_1, \Psi_2, \Psi_3$, respectively. We adopt the slowest libration period as the characteristic resonant timescale, $t^{\rm lib}_{\rm ref} \sim$ 50 yr.
\end{enumerate}

We compared the quadrupole forcing timescale ($t_{OC, g}^{\rm quad}$) for each draw of the posterior to $t^{\rm lib}_{\rm ref}$ as an estimate of how rapidly the outer companion's dynamical influence could disrupt the libration of the Laplace angles and thus the resonant stability of the inner system. We show this ratio in panel c) of Fig.~\ref{fig:theplot}, and see a qualitatively similar result to that shown in panel b) of Fig.~\ref{fig:theplot} using the coupling coefficient $\epsilon_{jf}$. We note that 7810/65,536 ($\sim 12\%$) of the posterior draws do not satisfy the criterion for the quadrupole approximation, 
\begin{equation}
    \epsilon_{oct} = \frac{a_g}{a_{\rm OC}} \frac{e_{\rm OC}}{1-e_{\rm OC}^2}
\end{equation}
where $\epsilon_{oct} > 0.1$ requires a higher-order octupole approach \citep[][their equation 29]{Naoz2016}, motivating the need for N-body simulations. Additionally, a limitation of this approach is that the posterior draws with $\rm{a_{\rm OC}} \lesssim$ 1 AU cannot be modelled as hierarchical given that $a_{\rm OC} \sim a_g$. Therefore, we turn to direct N-body simulations to evaluate the impact of these close-in orbital solutions of the outer companion on the inner system.

\subsection{N-Body Stability Analysis} \label{sec:nbody}

We performed a suite of N-body simulations of the TOI-178 system both with and without an outer companion. We used the \texttt{REBOUND} framework \citep{rebound1} to model the system and integrate the six inner planets' orbits as well as the outer companion's orbit on secular timescales. Specifically, we used the symplectic integrator \texttt{WHFast} \citep{wisdomholman, rein2015}, which is particularly suited to modelling systems with small deviations from Keplerian orbital motion. We set the integration time step to 0.09578 d, corresponding to $\sim 1/20$ of the period of the innermost planet in the system. Details of the simulation configuration and related tests can be found in Appendix \ref{apdx}.

We integrated each simulation for 60 forcing timescales ($t_{\rm quad}$) with inner and outer limits of $10^4$ and $10^6$ yr. The quadrupole forcing timescales of the outer companion on the inner system vary by eight orders of magnitude across the posterior ($10^{-1} - 10^7$ yr), corresponding to $10^0 - 10^8$ yr per integration if not capped. We set the maximum integration time to $10^6$ yr for computational tractability. As such, we do not claim that simulations that were undisrupted for $60\, t_{\rm{quad}}$ or $10^6$\,yr are stable, only that they are stable within the integration timescale. A simulation was labelled as unstable if any planet was ejected or had a close encounter with another planet in the system. We set the ejection criterion as three times the apocenter of the outer companion's orbit (3$a_{\rm OC}(1+e_{\rm OC})$). We set the close encounter criterion as the mutual Hill radius ($r_H$) of planets f and g, which is the largest $r_H$ in the inner system.

We initialize the simulations using a random draw of stellar and planet parameters for the inner TOI-178 system from the photodynamical chains reported in \citetalias{Leleu2024}. Specifically, we use the stellar mass, planet mass, semi-major axis, eccentricity, inclination, longitude of periapsis ($\varpi$), and mean longitude ($\lambda$) for each planet, except for TOI-178 b. This planet is not connected to the others in a resonant chain, and was fit with a circular orbit in \citetalias{Leleu2024}. We set the eccentricity of planet b to zero, and calculate its mean longitude of 0.1576 rad using its reported transit midpoint \citepalias{Leleu2024} assuming the longitude of the ascending node $\Omega_b=0$. We also include the effects of general relativity using the \texttt{gr$\_$potential} functionality of \texttt{REBOUNDx} \citep{nobili1986, reboundx}.

We then added an outer companion to the simulation. The giant companion was initialized using the semi-major axis, eccentricity, argument of periapsis ($\omega$), longitude of the ascending node ($\Omega$), inclination, reported epoch of periastron ($T_0$) from the G23H+RV posterior draws spanning the posterior parameter space. We did not run simulations for each posterior draw for computational tractability. Rather, we ran simulations for a subset of 1700 OC realizations, and present the results in panel d) of Fig. \ref{fig:theplot}. We find that the results of this analysis are consistent with those of the heuristic stability analyses reported in Section \ref{sec:stability}, namely that posterior draws within $\sim2$\,AU and which are not strictly hierarchical are more likely to lead to system instability over the integration timescale. We note that 260 simulations did not integrate for the full 60 $t_{\rm quad}$ and were instead capped at $10^6$ yr, and 134 simulations were integrated for more than 60 $t_{\rm{quad}}$ to reach the minimum integration time of $10^4$ yr. We found that 152 ($\sim 9\%$) of the simulations are unstable on the integration timescale, almost all (150/152) due to close encounters. We also found that the resonant chains in 172 of the simulations cease to librate over the integration, even if the simulation did not undergo a close encounter or ejection.

\section{Results} \label{sec:results}

We present the final reduced posterior sample for TOI-178\,h in Fig.~\ref{fig:final}. This posterior sample was generated from the \texttt{Octofitter} posterior resulting from a joint astrometry and RV fit. We excluded draws that were incompatible with AO imaging of the host star in Section \ref{sec:ao} (see panel (a) of Fig.~\ref{fig:theplot}). We then calculated the strength of the coupling between TOI-178 f and another inner planet relative to the perturbing influence of TOI-178\,h in Section \ref{sec:stab-eps} (see panel (b) of Fig.~\ref{fig:theplot}). We also calculated the forcing timescale of TOI-178\,h on planet g given relative to the longest libration timescale of the resonant chain in Section \ref{sec:stab-times} (see panel (c) of Fig.~\ref{fig:theplot}). Finally, we performed N-body simulations on a subset of posterior draws in order to capture the impact of relativistic apsidal precession of planet b, octopole-level pertubations on the inner system, and additional interactions between planets in this high-multiplicity system. We found that the majority of remaining draws remained stable for 60$t_{quad}$ (see panel (d) of Fig.~\ref{fig:theplot}), although this does not preclude later instabilities.

Using the results from these additional analyses, we exclude posterior draws from the G23H+RV fit which were incompatible with the AO contrast curve, fell in the weak-coupling regime ($\epsilon_{jf} > 10$), had quadrupole forcing times comparable to the libration of the resonant chain (${t_{g,OC}^{quad}}/{t_{ref}^{lib}} < 10$). This resulted in a final posterior sample of 44142 draws ($67\%$ of the initial posterior sample). As shown in Fig.~\ref{fig:final}, we constrain TOI-178\,h to have a semi-major axis of $6.9^{+12.3}_{-3.3}$\,AU and a mass of $21^{+47}_{-11}\,M_J$. As such, this OC is likely a brown dwarf with a mass ratio of $\mu = 0.03^{+0.07}_{-0.02}$ (or $\mu \sim 3\%$) at a moderate separation from TOI-178.

\section{Discussion}\label{sec:discussion}

\subsection{High-Multiplicity Resonant System Architectures}

While there are many small planets near MMR in multi-planet systems \citep[e.g.][]{Lissauer2011}, the sample of stars hosting resonant planet pairs is far smaller \citep[e.g.][]{turtelboom+22, Almenara2026} and there are only seven high-multiplicity ($N_{pl} > 4$) planetary systems hosting confirmed resonant chains of small planets (i.e. with three or more planets in resonance). K2-138 is the closest analogue to TOI-178, hosting six sub-Neptunes \citep{khu2021, christiansen18}, the inner five of which form a resonant chain \citep{Leleu2026}. Both HD 110067 \citep{Luque2023} and TOI-1136 \citep{Dai2023} host six-planet resonant chains of small planets. Kepler-80 \citep{MacDonald2021} and Kepler-223 both host close-in chains of four sub-Neptunes \citep{Mills2016}. Finally, the TRAPPIST-1 system hosts a chain of seven small planets within just 0.06 AU \citep{Papaloizou2018}. Notably, none of these systems' host stars have known stellar or substellar companions. Gaia DR4 (expected December 2026) has the potential to reveal OCs in these other high-multiplicity resonant chains. TOI-178 is currently the only known system that has maintained a high-multiplicity resonant chain whilst in the dynamical presence of a massive outer companion.


Resonant chains are thought to form in abundance in the first few Myr of system lifetimes as a consequence of convergent migration driven by parent gas disks \citep[e.g.][]{Izidoro2017}. However, most of these chains apparently disrupt after the disk has dispersed, on a characteristic timescale $\sim$100 Myr \citep{Dai2024}. The reason for the disruption is unclear \citep{Choksi2026, Hadden2026, Ogihara2026, LoRusso2026, Pichierri2020}. TOI 178 hosts one of the few mature systems that has managed to avoid instability. Its inner resonant chain and OC architecture encourages further investigation into how external secular effects may influence the dynamics and long-term stability of resonant chains \citep{Rodet2021, Choksi2023, Gu2026}.

\subsection{Outer Companion Formation \& Migration}

After dynamical considerations (see Section~\ref{sec:stability}), the majority of our refined orbital posterior for TOI-178\,h places its mass at $21^{+47}_{-11}\, M_J$ within the brown dwarf regime, suggesting that this object likely formed via direct collapse of protoplanetary disk material. Historically, brown dwarfs (BDs) were considered to be substellar objects with masses above the deuterium-burning limit ($\sim 13 M_J$) and below the hydrogen-burning limit ($\sim 80 M_J$). Previous observational works also identified the ``brown-dwarf desert'': a lack of objects with masses between $\sim 10-80\, M_J$ as companions to other stars \citep[e.g.][]{sahlmann2011}. More recently, this desert has been parametrised as a feature of mass-ratio rather than mass, with a lack of objects with mass ratios of a few percent \citep{duchene2023, zhang2025, cui2026, bddesert}.

Unlike giant planets that primarily form via core accretion \citep{Pollack1996}, BDs and very low mass stars (VLMs) are theorized to form via gravitational instability (GI) leading to direct collapse of disk material \citep{Boss1997, Stamatellos2009, Kratter2010}. GI can be triggered via disk cooling or a sufficient increase in the local gas surface density due to, for example, mass infall from the surrounding environment \citep{Longarini2025b}. Recent observational signatures have  lent additional evidence to giant planets and BDs/VLMs forming via distinct mechanisms. Namely, \citet{Bowler2020} found that BDs are typically more eccentric than cold giant planets. This is in line with TOI-178\,h, which appears to have an elevated eccentricity of $0.76^{+0.18}_{-0.35}$, but we caution that this constraint is drawn from data with incomplete orbital coverage for the majority of posterior draws. Additionally, \citet{Giacalone2026} reported that objects $>30\,M_J$ and beyond 1 AU typically orbit stars that are systematically less metal-rich than those with lower masses. This supports the formation of BDs via GI, which depends weakly on gas metallicity, contrary to core accretion that relies heavily on the availability of solids before significant gas accretion can be triggered \citep{fioscher2005, hasegawa2014, carter2026}. TOI-178 is known to be a metal-poor star \citep[$\rm{[Fe/H]}$ = -0.23 $\pm$ 0.05 dex][]{delrez2023}, which further disfavours the core-accretion scenario given that gas giants are known to be rare in metal-poor systems \citep{Gan2025}.

Simulations from \citet{Mercer2020} found that GI around low-mass stars (such as TOI-178) is able to produce giant companions ($\sim10\, M_J$) for initial disk-to-star mass ratios $>30\%$, with the majority of giant companions forming at $\sim$ 50 AU. The Disc Instability Population SYnthesis model \citep[DIPSY,][]{Schib2025} has also shown that BDs with masses of tens of Jupiter masses are the most common companions formed via GI around stars from A-type stars to late-type M dwarfs so long as disk fragments are not destroyed via collisions or ejections within the first $0.1$ Myr. However, these surviving BDs migrate inwards extremely rapidly on almost orbital timescales of $\sim10^3$ yr \citep{Baruteau2011}. GI is also a rapid process ($\sim 10^3$ yr) once the threshold for collapse is met \citep{Stamatellos2009}. This suggests that TOI-178\,h may have formed late in the protoplanetary disk's lifetime, after the inner planets formed and converged into a resonant chain. We speculate that the existence of TOI-178\,h may have stabilized the inner system's resonant chain from disruption once the disk dispersed, keeping the resonant chain intact. Future surveys, particularly from Gaia DR4 and DR5, will reveal whether other known resonant chain systems host similar OCs, thus providing constraints on how these unique planetary systems formed and evolved.

\begin{figure}[t!]
    \centering
    \includegraphics[width=\linewidth]{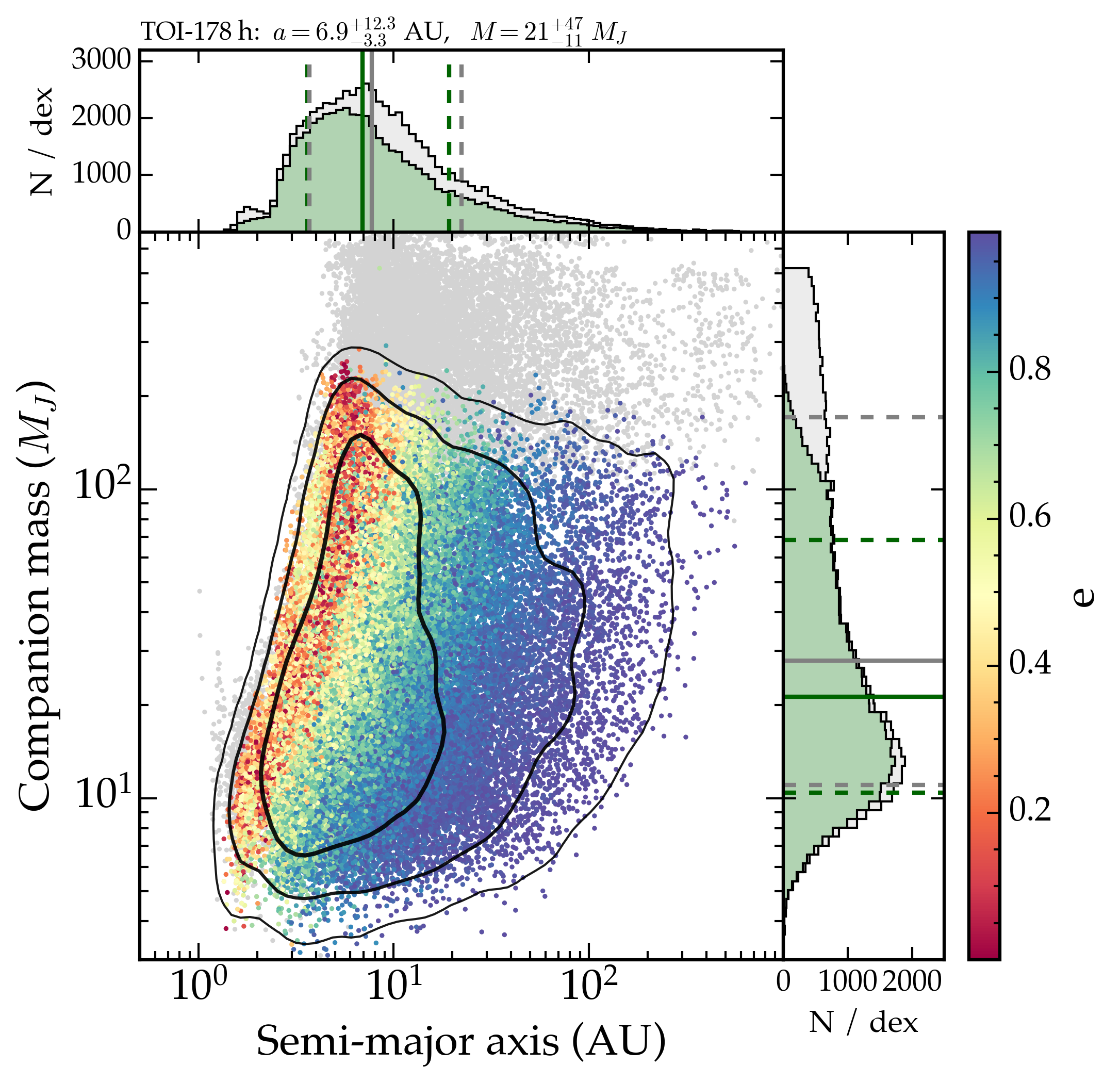}
    \caption{The final posterior draws for TOI-178\,h after excluding draws from the full G23H+RV posterior incompatible with AO imaging and those determined to be unstable via heuristic stability analyses or on the timescale of the N-body simulations. The full G23H+RV posterior draws are shown as grey points. The surviving posterior draws are shaded according to fitted eccentricity. The marginalised semi-major axis and mass posteriors are shown for the G23H+RV (grey) and final (green) posteriors in the top and right axes, with vertical lines showing the median (solid) and 16th/84th percentiles (dashed). KDE contours at 68.3$\%$, 95.4$\%$, and 99.7$\%$ are shown over the final posterior draws. }
    \label{fig:final}
\end{figure}

\section{Conclusions} \label{sec:conclusions}
In this paper, we presented the discovery of a massive outer companion to TOI-178, a K dwarf known to host six small transiting planets, five of which form a 2:4:6:9:12 resonant chain. We use calibrated, absolute astrometric observations from the G23H catalogue \citep{Thompson2026} alongside literature RVs from ESPRESSO  \citep{leleu2021} to reveal the presence of TOI-178\, h and constrain its orbit. We incorporate archival adaptive optics imaging from Keck-II/NIRC2 as well as both heuristic and N-body stability analyses to refine the mass--semi-major axis posterior distribution for this companion. We present our final posterior in Fig.~\ref{fig:final}, which suggests that TOI-178\,h is likely a brown dwarf with a semi-major axis of $6.9^{+12.3}_{-3.3}$\,AU and a mass of $21^{+47}_{-11}\,M_J$. TO1-178 the only confirmed system to host both a high-multiplicity resonant chain as well as a massive outer companion. We conclude that TOI-178\,h likely formed by direct collapse due to gravitational instabilities in the protoplanetary disk. The presence of TOI-178\,h raises the question of how the inner resonant chain was able to resist perturbation and breaking over the system's Gyr lifespan. 

\newpage

\section*{Data Availability}
The G23H posterior can be viewed and downloaded at \url{https://exopo.st/}.

\facilities{Keck/NIRC2, Gaia, Hipparcos}

\begin{acknowledgments}
RC acknowledges the support of the Natural Sciences and Engineering Research Council of Canada (NSERC) and the Canada Research Chairs (CRC) Program [Award ID: CRC-2024-00372]. DRC acknowledges partial support from NASA Grant 18-2XRP18$\_$2-0007. NC and JS acknowledge financial support from the Heising–Simons Foundation through the 51 Pegasi b Fellowship (grant \#2025-5885). DJ is supported by NRC Canada and by an NSERC Discovery Grant.

This research has made use of NASA’s Astrophysics Data System, and of the NASA Exoplanet Archive \citep{nea1}. This research has made use of the Exoplanet Follow-up Observation Program (ExoFOP; DOI: 10.26134/ExoFOP5) website, which is operated by the California Institute of Technology, under contract with NASA under the Exoplanet Exploration Program. Some of the data presented herein were obtained at Keck Observatory, which is a private 501(c)3 non-profit organization operated as a scientific partnership among the California Institute of Technology, the University of California, and NASA. The Observatory was made possible by the generous financial support of the W. M. Keck Foundation. The raw Keck Observatory data can be accessed from the Keck Observatory Archive, which is operated by the W. M. Keck Observatory and the NASA Exoplanet Science Institute (NExScI), under contract with NASA. The authors wish to recognize and acknowledge the very significant cultural role and reverence that the summit of Maunakea has always had within the Native Hawaiian community. We are most fortunate to have the opportunity to conduct observations from this mountain.

This work has made use of data from the European Space Agency (ESA) mission Gaia (\href{https://www.cosmos.esa.int/gaia}{https://www.cosmos.esa.int/gaia}), processed by the Gaia Data Processing and Analysis Consortium (DPAC),\\\href{https://www.cosmos.esa.int/web/gaia/dpac/consortium}{https://www.cosmos.esa.int/web/gaia/dpac/consortium.} 

The authors acknowledge the use of Anthropic’s Claude (Opus-5) model to assist with code debugging and figure formatting. This research was enabled in part by support provided by McMaster University and the Digital Research Alliance of Canada. We recognize that McMaster University is located on the traditional territories of the Mississauga and Haudenosaunee nations, and within the lands protected by the “Dish with One Spoon” wampum agreement.

\end{acknowledgments}

\clearpage

\appendix

\section{N-body Simulation Configurations} \label{apdx}

In order to determine the appropriate simulation configuration for the N-body tests described in Section \ref{sec:nbody}, we performed the following tests. We ran a pilot simulation for 10 randomly drawn companions for $10^3$\,yr. We calculated the relative energy error across the simulation as the Newtonian energy plus the corresponding GR potential. We measured the change in system energy, normalised by the total orbital energy of the inner planets, 
\begin{equation}
    \sum_{i} \frac{G M_* m_i}{2 a_i}
\end{equation}
so that the much massive companion's orbital energy did not dominate errors. Here $G$ is the gravitational constant, $M_*$ is the stellar mass, $m_i$ is the mass of a given inner planet, and $a_i$ is the semi-major axis of a given inner planet. We found that the energy error remained bounded in simulations with and without companions, and do not see evidence of numerical drift. 

We also tested setting the symplectic corrector order as 3 or 11 to evaluate whether the added complexity and thus time for the higher order was warranted for our simulation. We found that the relative energy error was practically unchanged. We also checked that the change in the inner planets' mean semi-major axes between the first and last $10\%$ of the simulation remained roughly constant in the absence of an outer companion, and therefore used corrector order 3 in our production simulations.

All integrations were run on Fir, a national high-performance computing cluster of the Digital Research Alliance of Canada. We measured a integration speed on the Fir cluster with our test runs of 0.057 s of wall time per simulated year. The cost of each run was then estimated as its integration length T times this rate plus 3 min for post-processing diagnostics. This is a conservative upper bound, as simulations that became unstable through ejection or close encounter ended early, and post-processing time was minimal. For the production runs, we ran each simulation serially on a single core.

\bibliography{refs}{}
\bibliographystyle{aasjournal}


\end{document}